\documentclass[pdftex,twocolumn,epjc3]{svjour3}          

\RequirePackage[T1]{fontenc}

\smartqed  
\usepackage{amssymb,amsfonts,amsmath,braket}

\RequirePackage{graphicx}
\RequirePackage{mathptmx}      
\RequirePackage{flushend}
\RequirePackage[numbers,sort&compress]{natbib}
\RequirePackage[colorlinks,citecolor=blue,urlcolor=blue,linkcolor=blue]{hyperref}
\usepackage{ latexsym}
\newcommand{{\bfkappa}}{\mbox{\boldmath${\kappa}$\unboldmath}}
\newcommand{{\bfg}}{\mbox{\boldmath$g$\unboldmath}}
\newcommand{{\bfa}}{\mbox{\boldmath$$\unboldmath}}
\newcommand{{\bfc}}{\mbox{\boldmath$c$\unboldmath}}
\newcommand{{\bfv}}{\mbox{\boldmath$v$\unboldmath}}
\newcommand{{\bfk}}{\mbox{\boldmath$k$\unboldmath}}
\newcommand{{\bff}}{\mbox{\boldmath$f$\unboldmath}}
\newcommand{{\bfF}}{\mbox{\boldmath$F$\unboldmath}}
\newcommand{{\bfA}}{\mbox{\boldmath$A$\unboldmath}}
\newcommand{\gradv}{\boldsymbol{\nabla}}
\def\v#1{{\bf#1}}

\usepackage{calligra}
\usepackage{calrsfs}

\journalname{Eur. Phys. J. C}

\begin{document}
\title{Comment on ``Monopole production via photon fusion and Drell-Yan processes:
MadGraph implementation and perturbativity via
velocity-dependent coupling and magnetic moment as novel
features''}

\author{Ricardo Heras}

\thankstext{}{e-mail: ricardo.heras.13@ucl.ac.uk}

\institute{Department of Physics and Astronomy, University College London, London, WC1E 6BT, UK
}

\date{Received: date / Accepted: date}

\maketitle

\begin{abstract}
\noindent In a recent study on monopole production [\href{https://doi.org/10.1140/epjc/s10052-018-6440-6}{Eur. Phys. J. C (2018) {\bf 78}: 966}], Baines et al added the potential of a magnetic dipole to the Wu-Yang potentials for the Dirac monopole and claimed that this modified Wu-Yang configuration does not affect the Dirac quantisation condition. In this comment, we argue that their claim is incorrect by showing that their modified Wu-Yang configuration leads to an infinite number of quantisation conditions. In their study, they also incorrectly identified the magnetic field of the monopole with the magnetic field of the Dirac string and its attached magnetic monopole.
\end{abstract}
In their recent paper on monopole production, Baines et al \cite{1} have made various conceptual and formal mistakes in connection with their idea of introducing a magnetic dipole in the Dirac theory of magnetic monopoles \cite{2,3}. Here, we wish to stress these mistakes. Our criticisms are restricted to the electromagnetic aspects of Dirac's theory of monopoles and the Wu-Yang configuration  \cite{4} for the Dirac monopole.
\vskip 5pt
1. The authors of the paper \cite{1}  claim that ``point-like monopoles, originally envisaged
by Dirac, are sources of singular magnetic fields for which the underlying theory, if any, is completely unknown.'' This is an incorrect statement. First of all, Dirac did not envisaged point-like magnetic monopoles but hypothetical nodal lines (semi-infinite magnetised lines with vanishing wave function) having the same end point where a magnetic monopole is assumed to exist. A quantum-mechanical argument on these nodal lines led him to his famous quantisation condition: $q g=n\hbar c/2$. Here, $q$ and $g$ denote electric and magnetic charges, $\hbar$ is the reduced Planck's constant, $c$ is the speed of light, $n$ represents an integer number, and we are adopting Gaussian units. Expressed by Dirac himself \cite{2}: ``Thus at the end point [of nodal lines] there will be a magnetic pole of strength [$g=n\hbar c/(2q)$].''

On the other hand, it is not true that the theory underlying point-like monopoles is ``completely unknown.'' On the contrary, the classical theory of point magnetic monopoles is well-known. It is described (in the static regime) by the equations $\gradv\cdot\v B_\texttt{mon}=4\pi g\delta(\v x)$ and $\gradv\times\v B_\texttt{mon}=0$. Furthermore, the classical theory of the Dirac string along the $z$ axis is also well-known \cite{5,6,7} and has been recently discussed in a review paper on the Dirac quantization condition \cite{8}. In the case of a string along $z>0,$ this theory is consistently described by the static equations \cite{6,7,8}: $\gradv\cdot \v B_\texttt{ms}\!=\!0$ and $\gradv\times\v B_\texttt{ms}= -4\pi g\Theta(z)\big[\delta(x)\delta'(y) \hat{\v x}-\delta'(x)\delta(y)\hat{\v y}\big]$, whose solution is given by
 \begin{align}
 \v B_\texttt{ms} =& \;\v B_\texttt{mon} +\v B_\texttt{string},\nonumber\\
                  = &\;\frac{g}{r^2}\hat{\v r}-  4\pi g\delta(x)\delta(y)\Theta(z)\hat{\v z},
\end{align}
where $\Theta(z)\!=\!0$ if $z<\!0$ and $\Theta(z)\!=\!1$ if $z\!>\!0$. In conclusion, the classical theories underlying magnetic monopoles and Dirac's strings are well-known.

\vskip 5pt
2. The authors of \cite{1} claim that ``due to the monopole's magnetic charge,
there is a magnetic field contribution, which however, due to the (singular) Dirac string, requires proper regularisation.'' They state that
\begin{align}
 \overrightarrow{B}^\texttt{reg}_\texttt{monopole} =& \;\overrightarrow{B}_\texttt{monopole} +\overrightarrow{B}_\texttt{sing},\nonumber\\
                  = &\;\frac{g}{r^2}\hat{\v r}-  4\pi g\hat{ n}\Theta(z)\delta(x)\delta(y),
\end{align}
is ``the regularised monopole magnetic field for a Dirac string along the $z$-axis, in which case the unit vector $\hat n = (0, 0, 1)$ also lies along that axis.'' First of all, it is the Dirac potential $-$and not its magnetic filed$-$ which must be regularised \cite{5,6,7,8}. Clearly, the right-hand sides of (1) and (2) are the same
\begin{align}
\v B_\texttt{ms}=\overrightarrow{B}^\texttt{reg}_\texttt{monopole},
\end{align}
and therefore $\overrightarrow{B}^\texttt{reg}_\texttt{monopole}$  represents the magnetic field of the Dirac string with its attached magnetic monopole. Nevertheless, they claim that $\overrightarrow{B}^\texttt{reg}_\texttt{monopole}$ describes the ``regularised'' magnetic field of the monopole and therefore they point out that this field ``yields the correct formula
\begin{align}
\gradv\cdot\overrightarrow{B}^\texttt{reg}_\texttt{monopole}=4\pi g\delta^3(\v r),
\end{align}
implying that the magnetic monopole is the source of a field.'' However, the formula (4) is wrong. The correct formula is
\begin{align}
\gradv\cdot\overrightarrow{B}^\texttt{reg}_\texttt{monopole}=0,
\end{align}
which follows from taken the divergence to (2) (see \cite{6,7,8}),
\begin{align}
\gradv\cdot\overrightarrow{B}^\texttt{reg}_\texttt{monopole}=&\;\gradv\cdot\overrightarrow{B}_\texttt{monopole} +\gradv\cdot\overrightarrow{B}_\texttt{sing}\nonumber\\
=&\;4\pi g\delta^3(\v r)-4\pi g\delta^3(\v r).
\end{align}
In other words, they incorrectly identify the field $ \overrightarrow{B}^\texttt{reg}_\texttt{monopole}$ with the magnetic field of a magnetic monopole. From (3) it follows that the field
$\overrightarrow{B}^\texttt{reg}_\texttt{monopole}$ must be identified with the field of the Dirac string with its attached magnetic monopole.
\vskip 5pt
3. The authors of \cite{1} place a point dipole possessing the magnetic dipole moment  $\overrightarrow{\mu_\texttt{D}}$ at the origin of a coordinate system, where the monopole (attached to the Dirac string) is assumed be at rest. They note that the magnetic field produced by this dipole is given by
\begin{align}
 \overrightarrow{B}_\texttt{D} = &\;\frac{\mu_0}{4\pi r^3}|\overrightarrow{\mu_\texttt{D}}|\big(2 \cos{\theta}\hat{\v r}+\sin{\theta}\hat{\theta}\big),
\end{align}
(they work in spherical coordinates and tacitly assume that the magnetic dipole moment is parallel to the $z$-axis). They claim that ``a magnetic dipole moment does not contribute to the singular part of the magnetic field of the monopole, which is responsible for the charge quantisation.'' To support their claim they elaborate two arguments.

According to their first argument, the charge $q$ that encircles the Dirac string  must be placed ``far away from the position of the monopole'' and consequently the effects of the monopole are practically ignorable. Under this assumption, we can see that  only the singular string magnetic field  $ \overrightarrow{B}_\texttt{sing}= -  4\pi g\hat{n}\Theta(z)\delta(x)\delta(y)$ contributes to the phase of the wave function of the charge.  As is well-known, this singular field leads to the Dirac quantisation condition, which explains the quantisation of the electric charge. Since the magnetic dipole is placed at the same point that the magnetic monopole (the origin of coordinates) the authors assume that the effects of the former are also ignorable and conclude that ``The magnetic dipole moment does not contribute to the singular part of the magnetic field [$ \overrightarrow{B}_\texttt{sing}= -  4\pi g\hat{ n}\Theta(z)\delta(x)\delta(y)$], and thus the charge quantisation is not affected.'' This argument is clearly unsatisfactory. What exactly means far away from the position of the monopole? According to the configuration proposed by these authors, the Dirac quantisation condition $-$and hence the explanation of charge quantisation$-$ holds only far away from the position of the monopole and the dipole. This result disagrees with the idea that the quantisation of the electric charge is a universal law which does not depend on the point where we are observing the charge. We really find unsatisfactory the idea implied by the assumption of the authors that
when approaching to the origin (where the monopole and dipole are located), the explanation of charge quantisation provided by the Dirac quantisation condition is no larger valid.

In their second argument, the authors invoked the well-known Wu-Yang potentials  \cite{4}. They  wrote the following paragraph: ``For the magnetic monopole gauge potential one has the expression
\begin{align}
\overrightarrow{A}^S=& \,\,g\frac{1-\cos\theta}{r \sin\theta}\hat{\phi},\qquad\quad \theta\in \bigg[0, \frac{\pi}{2}+\delta\bigg)\quad\delta\to 0^+
\end{align}
for the south hemisphere, which is singular at the south pole $\theta=\pi$, and
\begin{align}
\overrightarrow{ A}^N=&  -g\frac{1+\cos\theta}{r \sin\theta}\hat{\phi},\quad \quad \theta\in \bigg(\frac{\pi}{2}-\delta, \pi\bigg]\quad\delta\to 0^+
\end{align}
for the north hemisphere, which is singular at the north pole $\theta=0$. These two patches overlap $\pi/2-\delta<\theta<\pi/2+\delta$, $\delta\to 0^+$
, and, as is well known, the difference of
\begin{align}
\overrightarrow{A}^S-\overrightarrow{A}^N=\gradv f=\frac{2 g}{r\sin\theta}\hat{\phi},
\end{align}
yields a singular gauge transformation at $\theta=0,\pi,$ which contributes to the phase $q_e\oint_{L}d \overrightarrow{x}\cdot \overrightarrow{A}$ of the charged particle wavefunction, the requirement of single-valuedness of which yields the rule [$qg = n \hbar c/2$].'' This paragraph contains two mistakes. First, they confused the North Wu-Yang potential with the South Wu-Yang potential (see \cite{5,8}). Second, they wrote (10) without specifying that the factor $\sin \theta$ necessarily takes the value $1$ as $\delta\to 0^+$ in the overlapped region. The correct form is $\overrightarrow{A}^S-\overrightarrow{A}^N=\gradv f=2 g\hat{\phi}/r,$ which is non-singular in the overlapped region. Apparently, the form of (10) led them to incorrectly state that this is a singular gauge transformation at the angles $\theta=0,\pi$. In the Wu-Yang approach, these angles (and hence the corresponding Dirac strings) are excluded. As is well-known, the advantage of the Wu-Yang approach is precisely to avoid a singular gauge transformation!

Let us write the above paragraph of the authors in a correct way \cite{8}. Following the Wu-Yang method \cite{4}, the well-known Dirac potentials $\v A'= g(1-\cos\theta)\hat{\phi} /(r \sin\theta)$ and $\v A= -g(1+\cos\theta)\hat{\phi}/(r \sin\theta)$ are non-singular if we define them in an appropriate domain:
\begin{align}
\v A'=& \,\,g\frac{1-\cos\theta}{r \sin\theta}\hat{\phi},\qquad\quad R^N:\;0 \leq \theta < \frac{\pi}{2}+\frac\varepsilon2\\
\v A =&  -g\frac{1+\cos\theta}{r \sin\theta}\hat{\phi},\quad \quad R^S:\;\frac{\pi}{2}-\frac\varepsilon2 < \theta \leq \pi
\end{align}
where $\varepsilon>0$ is an infinitesimal quantity. The potentials  $\v A'$ and $\v A$ are in the Coulomb gauge: $\gradv \cdot \v A =0$ and $\gradv \cdot \v A'=0$. Furthermore, these potentials are non-global functions since they are defined only on their respective domains: $R^N$ and $R^S$. The region $R^N$, where $\v A'$ is defined, excludes the string along the negative semi-axis $(\theta=\pi)$ and represents a North hemisphere. The region $R^S$, where $\v A$ is defined, excludes the string along the positive semi-axis $(\theta=0)$ and represents a South hemisphere. The union of the  hemispheres $R^N\cup R^S$ covers the whole space (except on the origin, where the monopole is located). In the intersection region $R^N\cap R^S$  (the ``equator'') both hemispheres are slightly overlapped. Each one of the potentials $\v A'$ and $\v A$ yield the field of a magnetic monopole: $\v B = \gradv \times \v A'= \gradv \times \v A= g \hat{\v r}/r^2$. Therefore, the Coulomb-gauge potentials  $\v A'$ and $\v A$ must be connected by a $restricted$ gauge transformation in the equator, i.e., in the overlapped region $\pi/2-\varepsilon/2 < \theta <\pi/2+\varepsilon/2$, where both potentials are well defined. At first glance, we would have $\v A'-\v A=2 g\hat{\phi}/(r \sin\theta)$. But in the overlapped region, we have $\lim \sin(\pi/2\pm\varepsilon/2)=1$ as $\varepsilon\to 0$ and therefore
\begin{align}
\v A'-\v A=\frac{2 g}{r}\hat{\phi}=\gradv (2g\phi)=\gradv \Lambda,
\end{align}
where $\Lambda =2 g \phi$ is a multi-valued gauge function [$\Lambda(\phi)\neq \Lambda(\phi+2\pi)$] satisfying the Laplace equation $\gradv^2\Lambda=0.$ Unfortunately, it is not usually emphasised in the literature $-$but it should be$-$ that the gauge function $\Lambda$ in (13) connecting the Coulomb-gauge potentials $\v A'$ and $\v A$ corresponds to a restricted gauge transformation \cite{8}. If an electric charge $q$ is interacts with the monopole we require two wave functions to describe the electric charge: $\psi'$ for $ R^N$ and $\psi$ for $ R^S$. In the equator, the wave functions $\psi'$ and $\psi$ must be connected by the phase transformation $\psi'={\rm e}^{iq\Lambda/(\hbar c)}\,\psi$, which is related to the gauge transformation (13). Substituting $\Lambda =2 g \phi$ in the phase transformation, we obtain the multi-valued wave function $\psi'={\rm e}^{i2qg \phi/(\hbar c)}\,\psi,$ i.e., $\psi'|_\phi\not=\psi'|_{(\phi+2\pi)}$. But we require the wave function be single-valued, i.e.,
 $\psi'|_\phi=\psi'|_{(\phi+2\pi)}$ and therefore we require ${\rm e}^{i4 \pi qg/ (\hbar c )}\!=\!1$ and this implies the Dirac quantisation condition $qg=n\hbar c/2$.
\vskip 5pt
4. On the basis of their considerations on the Wu-Yang potentials, the authors of \cite{1} ``present an
equivalent, yet less elaborate, way to see the irrelevance of
the magnetic dipole moment for the quantisation rule  $[qg=n\hbar c/2$],
which avoids the use of Dirac strings.'' With this aim they state that ``the vector potential
corresponding to the magnetic moment, for large distances
$r$ from the centre of the sphere where the monopole is located, is of the form
\begin{align}
\overrightarrow{A}_\texttt{D}=\frac{\mu_0}{4\pi}\frac{\overrightarrow{\mu}_\texttt{D}\times \overrightarrow{ r}}{r^3}=\frac{\mu_0}{4\pi}\frac{|\overrightarrow{\mu}_\texttt{D}|\sin\theta}{r^2}\hat{\eta},
\end{align}
with $\hat{\eta}$ the unit vector perpendicular to the plane of $\overrightarrow{r}$ and $\overrightarrow{\mu}_\texttt{D}$
(assumed parallel to the $z$-axis); this is not singular at the poles $\theta=0,\pi$ (in fact it vanishes there). The total potential
in each hemisphere is then given by the corresponding sum
$\overrightarrow{A}_i + \overrightarrow{A}_D, i = S, N.$ Hence, the magnetic moment does
not contribute to the difference, and thus it does not affect the wave function phase, which is associated only with the
monopole part [(10) of this comment].'' We will show that this conclusion is wrong.

Let us consider the idea of the authors to add the potential of the magnetic dipole $\v A_\texttt{D}$
to the South Wu-Yang potential (12), for example. Of course, we could equally consider the North Wu-Yang potential (11). For consistence we adopt Gaussian units in which the potential of the dipole reads
\begin{align}
\v A_{\texttt{D}}=\frac{|\v \mu |\sin\theta}{cr^2}\hat{\phi}.
\end{align}
(notice that the unit vector $\hat{\eta}$ introduced by the authors in (14) is the same as $\hat{\phi}$ because the magnetic moment is assumed to be parallel to the $z$-axis). Thus, the ``total potential" in the South hemisphere is given by
\begin{align}
\v A_{\texttt{total}} =\v A + \v A_\texttt{D}.
\end{align}
This potential has azimuthal symmetry, i.e, $\v A_{\texttt{total}}= A_{\texttt{total}}\hat{\phi}$ and is in the Coulomb gauge $\gradv\cdot \v A_{\texttt{total}}= 0$ because $\gradv\cdot \v A= 0$ and $\gradv\cdot \v A_{\texttt{D}}= 0.$ The subtle point here is that the South Wu-Yang potential $\v A$  is already connected with the North Wu-Yang potential $\v A'$ by means of the restricted gauge transformation (13) which holds in the equator. Accordingly, there is no arbitrariness in the potential $\v A$ (or in the potential $\v A'$ for the same reason). But we cannot say the same for the potential of the magnetic dipole $\v A_\texttt{D}$ which can be subject of a further restricted gauge transformation $\v A'_{\texttt{D}}=\v A_{\texttt{D}}+\gradv\xi$, where $\gradv^2\xi=0$. Thus there is still arbitrariness in the potential  $\v A_{\texttt{D}}$ which can be translated to the total potential $\v A_{\texttt{total}}.$ Put differently, we are free to apply a restricted gauge transformation to the total potential, i.e., $\v A'_{\texttt{total}} = \v A_{\texttt{total}}+ \gradv\chi$ where $\chi$ is an arbitrary gauge function to the extent that it satisfies $\gradv^2\chi=0.$ In fact, in the equator (where the Wu-Yang potentials $\v A$ and $\v A'$ are well-defined) the gauge function $\chi$ may take the generic form $\chi= 2g  \phi + \xi.$ The first term is the gauge function $\Lambda=2 g \phi$ of the gauge transformation (13) that relates the Wu-Yang potentials and the second term is an arbitrary gauge function $\xi$ connecting the two Coloumb-gauge dipole potentials $\v A_{\texttt{D}}$ and $\v A'_{\texttt{D}}.$ It is the arbitrariness of $\xi$ that makes $\chi$ arbitrary and consequently $\v A_\texttt{D}$ is arbitrary up to a restricted gauge transformation. In particular, we are free to choose $\xi=k \phi,$ where $k$ is an arbitrary constant with units of magnetic charge. With this choice for $\xi$ we can consider the restricted gauge transformation at the equator (where $\v A$ and $\v A'$ are well-defined)
\begin{align}
\v A'_{\texttt{total}}-\v A_{\texttt{total}} = \frac{2 g + k}{r}\hat{\phi}=\gradv (2g\phi+ k\phi) = \gradv \chi,
\end{align}
where $\chi= (2g +  k)\phi$ is a multi-valued gauge function satisfying $\gradv^2 \chi=0.$ Following the Wu-Yang approach \cite{4}, the wave functions $\psi'$ and $\psi$ must be connected by the phase transformation $\psi'={\rm e}^{iq\chi/(\hbar c)}\,\psi$ in the equator. Substituting $\chi =(2g +k)\phi$ in the phase transformation we obtain the multi-valued wave function $\psi'={\rm e}^{iq(2g +k)\phi/(\hbar c)}\,\psi$. The condition of single-valuedness of the wave function \big($\psi'|_\phi=\psi'|_{(\phi+2\pi)}$\big) requires ${\rm e}^{i2 \pi q (2g +k)/ (\hbar c )}\!=\!1$ and this implies the quantisation condition
\begin{align}
q \bigg(g +\frac{k}{2}\bigg) = \frac{n}{2}\hbar c.
\end{align}
This expression represents an infinite number of quantisation conditions because of the arbitrariness of the constant $k$.
Therefore, the idea of the authors of the paper \cite{1} that the addition of the potential of a magnetic dipole to the Wu-Yang potentials does not affect the Dirac quantisation condition is incorrect.

\end{document}